\documentclass[aps,pre,superscriptaddress,twocolumn,a4paper,reprint,noeprint]{revtex4-1}
\usepackage[utf8]{inputenc}
\usepackage{graphicx}
\usepackage{amsmath}
\usepackage{amsfonts}
\usepackage{amssymb}
\usepackage{bbold}
\usepackage[usenames,dvipsnames,svgnames,table]{xcolor}
\usepackage[USenglish]{babel}
\usepackage[colorlinks,linkcolor=blue,urlcolor=blue,citecolor=blue]{hyperref}
\usepackage{grffile}
\usepackage[normalem]{ulem}

\newcommand{\air}{\rule[-1mm]{0mm}{4mm}}
\newcommand{\moreair}{\rule[-3mm]{0mm}{8mm}}
\newcommand{\ignore}[1]{}
\newcommand{\E}[1]{{}^#1\!E}

\allowdisplaybreaks
\begin{document}

\title{Ginzburg--Landau theory for unconventional surface superconductivity in PtBi\textsubscript{2}}

\author{Harald Waje}
\affiliation{Institute of Theoretical Physics, Technische Universit\"at Dresden, 01062 Dresden, Germany}
\author{Fabian Jakubczyk}
\affiliation{Institute of Theoretical Physics, Technische Universit\"at Dresden, 01062 Dresden, Germany}
\affiliation{W\"urzburg-Dresden Cluster of Excellence ct.qmat, Technische Universit\"at Dresden, 01062~Dresden, Germany}
\author{Jeroen van den Brink}
\affiliation{Institute of Theoretical Physics, Technische Universit\"at Dresden, 01062 Dresden, Germany}
\affiliation{W\"urzburg-Dresden Cluster of Excellence ct.qmat, Technische Universit\"at Dresden, 01062~Dresden, Germany}
\affiliation{Leibniz Institute for Solid State and Materials Research (IFW) Dresden, Helmholtzstrasse 20, 01069~Dresden, Germany}
\author{Carsten Timm}
\email{carsten.timm@tu-dresden.de}
\affiliation{Institute of Theoretical Physics, Technische Universit\"at Dresden, 01062 Dresden, Germany}
\affiliation{W\"urzburg-Dresden Cluster of Excellence ct.qmat, Technische Universit\"at Dresden, 01062~Dresden, Germany}

\date{July 3, 2025}

\begin{abstract}
Recent experimental evidence suggests the presence of an unconventional, nodal surface-su\-per\-con\-duc\-ting state in trigonal PtBi\textsubscript{2}. We construct a Ginzburg--Landau theory for the three superconducting order parameters, which correspond to the three irreducible representations of the point group $C_{3v}$. The irreducible representations $A_1$ and $A_2$ are the most likely. We develop a systematic method to determine the symmetry-allowed terms and apply it to derive all terms up to fourth order in the three order parameters. The Ginzburg--Landau functional also includes coupling to the magnetic field. The functional is employed to determine the effect of an applied uniform magnetic field on the nodal structure for $A_1$ and $A_2$ pairing. The results facilitate clear-cut experimental differentiation between these symmetries. We also predict field-induced helical superconductivity.
\end{abstract}

\maketitle

\section{Introduction}

Trigonal PtBi\textsubscript{2} has recently emerged as a rather puzzling unconventional superconductor. The crystal structure with the noncentrosymmetric space group $P31m$ \cite{KBR14, SKP20} and point group $C_{3v}$ consists of Bi-Pt-Bi triple layers, where the two Bi layers are inequivalent. Consequently, there are two inequivalent $(001)$ surfaces, a corrugated ``decorated honeycomb'' Bi layer and a planar ``kagome'' Bi layer, respectively. The point group $C_{3v}$ is not reduced by the presence of these surfaces.

Density functional theory (DFT) calculations \cite{VLB23, KSV24, VKF24, VKV24} predict 12 Weyl points in general momentum positions. Since all Weyl points are related by point-group or time-reversal symmetries they are at the same energy, about $45\,\mathrm{meV}$ above the Fermi energy \cite{VKV24}. They are located close to the $\Gamma$MLA mirror planes \cite{endnote.mirrors}. Due to the presence of Weyl points, one expects six Fermi arcs on $(001)$ surfaces connecting the projections of the Weyl points into the surface (or slab) Brillouin zone. DFT calculations for slabs indeed find these arcs \cite{KSV24, VKF24}. They are different for the two Bi terminations but in both cases are horseshoe shaped and connect neighboring Weyl points across the $\Gamma$M lines. Quasi-particle interference \cite{HSV24} and angle-resolved photoemission spectroscopy (ARPES) \cite{KSV24, CSK24} in the normal state provide strong evidence for the Fermi arcs at both surfaces.

The superconducting properties of trigonal PtBi\textsubscript{2} are puzzling. Initially, single crystals were found to become superconducting at a broad transition at about $T_c = 600\,\mathrm{mK}$ \cite{SKP20}. More recent measurements on high-quality single crystals showed a much sharper transition at $T_c = 1.1\,\mathrm{K}$ \cite{ZCC24}. Wang \textit{et al.}\ \cite{WCZ21} only observed superconductivity under pressure, with a critical pressure of $5$ to $6\,\mathrm{GPa}$ and an otherwise weakly pressure-dependent critical temperature of about $2\,\mathrm{K}$~\cite{WCZ21}.

Point-contact measurements \cite{BKS22} showed enhanced $T_c \approx 3.5\,\mathrm{K}$, attributed to the higher local density of states in the vicinity of the point contact. The authors also concluded that electron-phonon coupling is a plausible mechanism.

Transport measurements by Veyrat \textit{et al.}\ \cite{VLB23} on thin exfoliated flakes showed a superconducting $T_c$ of a few hundreds of mK. For a thickness of $60\,\mathrm{nm}$, the authors found $T_c = 370\,\mathrm{mK}$. The current-voltage characteristics exhibited power-law scaling $V \sim I^a$ over a moderate range of currents, interpreted in terms of a Berezinskii--Kosterlitz--Thouless (BKT) transition \cite{Ber72, KoT73a, KoT73b} with critical temperature $T_\mathrm{BKT} = 310\,\mathrm{mK}$ determined by $a(T_\mathrm{BKT}) = 3$ \cite{VLB23}. This indicates two-dimensional (2D) superconductivity in spite of the film being much thicker than other systems showing BKT scaling \cite{VLB23}. Moreover, the critical magnetic field as a function of its angle relative to the surface exhibits a cusp-like maximum for the field in plane \cite{VLB23}, consistent with the Tinkham model for 2D superconductivity~\cite{Tin63a, Tin63b}.

The Fermi-arc signatures in laser ARPES exhibit shifts as functions of temperature that are characteristic for superconducting transitions \cite{KSV24}. The critical temperatures for superconductivity in the Fermi arcs are estimated as $T_c = 14 \pm 2\,\mathrm{K}$ for the decorated honeycomb surface and $T_c = 8 \pm 2\,\mathrm{K}$ for the kagome surface. The corresponding superconducting gaps are $1.4 \pm 0.2\,\mathrm{meV}$ and $2.0 \pm 0.2\,\mathrm{meV}$, respectively. The intensity vs.\ energy for the arc states shows an extremely sharp peak below $T_c$, interpreted as a superconducting coherence peak \cite{KSV24}. There is no indication of superconductivity in the bulk down to $1.5\,\mathrm{K}$.

Very recent ARPES experiments \cite{CSK24} with improved resolution have brought a big surprise: They show that the superconducting gap vanishes, within experimental accuracy, at the arc centers, where the arcs cross the $\Gamma$M lines. The gap appears to open linearly as a function of momentum along the arcs. While ARPES is not sensitive to the sign, or more generally the phase, of the gap, the results are most naturally explained by point nodes with linear dispersion and a sign change of the pairing amplitude \cite{CSK24}. It seems unlikely that a gap minimum without sign change shows fine-tuned linear dispersion. The maximum gap is on the order of $3\,\mathrm{meV}$. The critical temperature lies between $10$ and $15\,\mathrm{K}$~\cite{CSK24}.

Recent scanning-tunneling spectroscopy (STS) experiments \cite{SFH24} show tunneling gaps with characteristic coherence peaks at a temperature of $5\,\mathrm{K}$ for both surfaces. Although the surface topography indicates high surface quality with few point defects, the gap is spatially highly nonuniform spanning a range from $0$ to $20\,\mathrm{meV}$.

Scanning SQUID measurements \cite{SFH24} show a weak diamagnetic response at $6.4\,\mathrm{K}$, characteristic for 2D superconductivity. However, vortices are not observed. No anomaly related to superconductivity is observed in specific-heat measurements, which is consistent with the absence of bulk superconductivity. Surface superconductivity in PtBi\textsubscript{2} accompanied by a normal bulk is possible since the density of states at the surface is higher due to the Fermi arcs and can result in a much higher mean-field transition temperature for the surface \cite{NoH23, TKF24}.

In the absence of a good understanding of the microscopic mechanism, it is worthwhile to perform an analysis that only relies on symmetries. In this work, we set up a Ginzburg--Landau (GL) theory for possible superconducting states at the surface of PtBi\textsubscript{2}. The rather low symmetry described by the point group $C_{3v}$ and the correspondingly small number of irreducible representations (irreps) allows us to construct an essentially complete GL functional involving all possible couplings between the order parameters (OPs), also including gradients and the applied magnetic field.

The remainder of this paper is organized as follows. In Sec.\ \ref{sec.nodal}, we review the possible superconducting states and their symmetry-imposed nodal structure. In Sec.\ \ref{sec.GL}, we construct the GL functional and employ it to predict the change of the nodal structure when a magnetic field is applied. Finally, in Sec.\ \ref{sec.summary}, we summarize our work and draw conclusions.

\section{Symmetry and nodal structure of superconducting states}
\label{sec.nodal}

In this section, we review the symmetry analysis and resulting nodal structure of surface superconductivity in PtBi\textsubscript{2} \cite{CSK24} and provide additional details. The point group $C_{3v}$ is of order $6$ and contains the identity $\epsilon$, a rotation $C_3$ by $120^\circ$, its square $C_3^2$, and three vertical mirror planes, one of which is denoted by $\sigma_v$. It has the three irreps $A_1$, $A_2$, and $E$. Table \ref{tab.char} shows the characters and low-order basis functions of these irreps \cite{DDJ07, Katzer, endnote.mirrors}. The lowest-order basis function of $A_2$ is $x(y+x/\sqrt{3})(y-x/\sqrt{3})$, of order $l=3$ \cite{Katzer, endnote.mirrors}. To be able to safely construct terms in the GL functional, we have to carefully assign the first and second components of the 2D irrep $E$, which we denote by $\E1$ and $\E2$, respectively. (Note that most tables of basis functions do not distinguish between the components.) We take $(x,y)$ as the template, i.e., we require the first component to be odd under the mirror reflection $x \mapsto -x$ and the rotation $C_3$ to map the components as
\begin{align}
x &\stackrel{C_3}{\longmapsto} -\frac{1}{2}\, x + \frac{\sqrt{3}}{2}\, y , \\
y &\stackrel{C_3}{\longmapsto} -\frac{1}{2}\, y - \frac{\sqrt{3}}{2}\, x .
\end{align}
It can be checked that the other doublets given for the irrep $E$ in Table \ref{tab.char} transform in the same manner.

\begin{table}
\begin{center}
\caption{\label{tab.char}Characters and basis functions up to the order $l=2$ of the point group $C_{3v}$ \cite{DDJ07, Katzer, endnote.mirrors}. The irreps corresponding to angular-momentum or magnetic-field components $R_x$, $R_y$, $R_z$ are also shown.}
\begin{tabular}{ccccl} \hline\hline
\air & $\epsilon$ & $2C_3$ & $3\sigma_v$ & Basis functions \\ \hline
\air $A_1$ & $1$ & $1$ & $1$ & $1$, $z$, $x^2+y^2$, $z^2$ \\
\air $A_2$ & $1$ & $1$ & $-1$ & $R_z$ \\
\moreair $E$ & $2$ & $-1$ & $0$ & $\begin{pmatrix} x \\ y \end{pmatrix}$,
  $\begin{pmatrix} xy \\ (x^2-y^2)/2 \end{pmatrix}$,
  $\begin{pmatrix} xz \\ yz \end{pmatrix}$,
  $\begin{pmatrix} R_y \\ -R_x \end{pmatrix}$ \\ \hline \hline
\end{tabular}
\end{center}
\end{table}

We consider 2D superconductivity at a single surface of PtBi\textsubscript{2}, in practice realized by making the bulk sufficiently thick to decouple the surfaces. Our analysis applies to both the decorated honeycomb and the kagome termination but the values of parameters will of course be different. There are three possible complex superconducting OPs, corresponding to the irreps $A_1$, $A_2$, and $E$. Since $E$ is a 2D irrep, the corresponding OP generally has two complex components.

The position of nodes can be obtained based on the symmetry of the superconducting state \cite{CSK24}. The normal-state Fermi arcs are nondegenerate. Hence, at low energies, the superconducting state can be described by projecting a multiband Bogoliubov--de Gennes (BdG) Ha\-mil\-to\-nian
\begin{equation}
\mathcal{H}(\vec k) = \begin{pmatrix}
  \hat H_N(\vec k) & \hat\Delta(\vec k) \\
  \hat\Delta^\dagger(\vec k) & -\hat H_N^T(-\vec k)
\end{pmatrix} ,
\label{eq.BdG.2}
\end{equation}
where the block $\hat H_N(\vec k)$ is the normal-state Hamiltonian and $\hat\Delta(\vec k)$ is a pairing matrix, onto this single band. This relies on the energy splitting between the Fermi-arc band and bulk bands being larger than the superconducting energy scale. This condition breaks down in the vicinity of the Weyl points but ARPES experiments \cite{KSV24, CSK24} show that it holds over nearly the full arc.

The projected BdG Hamiltonian has the generic form
\begin{equation}
\mathcal{H}_\mathrm{proj}(\vec k) = \begin{pmatrix}
  \xi(\vec k) & \Delta(\vec k) \\
  \Delta^\ast(\vec k) & -\xi(-\vec k)
\end{pmatrix} ,
\label{eq.BdG.3}
\end{equation}
where $\xi(\vec k)$ is the normal-state dispersion and $\Delta(\vec k)$ is the pairing amplitude. Since there is no sign of time-reversal symmetry being broken spontaneously $\Delta(\vec k)$ can be chosen real for all momenta $\vec k$. For our analysis, we only require the dependence of the pairing amplitude along the Fermi arcs. To parametrize $\Delta$, we start from the polar angle $\phi$ of $\vec k$. $\phi$ does not uniquely label the arc states due to the horseshoe shape of the Fermi arcs. We can, however, remove this problem by deforming the parametrization without changing the symmetry \cite{CSK24}. The Fermi arcs are connected to short sections of bulk Fermi surfaces in the vicinity of the Weyl points that link them to the Fermi arcs at the opposite surface. The OP is, so far, unmeasurably small in the bulk~\cite{KSV24, CSK24}. For large intervals of angles $\phi$, there is no normal-state Fermi surface.

For every irrep $A_1$, $A_2$ or irrep component $\E1$, $\E2$, the function $\Delta(\phi)$ must be a real-valued basis function of this irrep or component. In Table \ref{tab.fphi}, we list the basis functions of the angle $\phi$ up to order $l=6$. Moreover, there is another important symmetry constraint: Fermionic antisymmetry requires the pairing matrix in the full BdG Hamiltonian, Eq.\ (\ref{eq.BdG.2}), to satisfy $\hat\Delta^T(-\vec k) = -\hat\Delta(\vec k)$ \cite{Gor58, SiU91, CTS16}. Simply speaking, this implies that the pairing matrix must be even under time reversal \cite{TiB21}. Time reversal maps $\vec k$ to $-\vec k$ and thus $\phi$ to $\phi + \pi$ modulo $2\pi$. The pairing amplitude $\Delta(\phi)$ in the projected BdG Hamiltonian, Eq.\ (\ref{eq.BdG.3}), inherits this property. The corresponding signs under time reversal are also given in Table~\ref{tab.fphi}.
Hence, the pairing amplitude $\Delta(\phi)$ must be a real basis function of the irrep or irrep component and must be even under time reversal. To find the symmetry-imposed nodes, it is sufficient to consider the lowest-order basis function with these properties. Basis functions of higher order only introduce additional modulations of the amplitude without changing its symmetry. More details are given in Appendix~\ref{app.symmetries}.

\begin{table}
\begin{center}
\caption{\label{tab.fphi}Low-order basis functions of the angle $\phi$, its order $l$, corresponding irreps, and signs under time reversal $\mathcal{T}$.}
\begin{tabular}{c@{\quad}c@{\quad}c@{\quad}c} \hline\hline
\air Basis function & Order $l$ & Irrep & Sign under $\mathcal{T}$ \\ \hline
\air $1$ & 0 & $A_1$ & $+$ \\
\moreair $\begin{pmatrix} \cos\phi \\ \sin\phi \end{pmatrix}$ & 1 & $E$ & $-$ \\
\moreair $\begin{pmatrix} \sin2\phi \\ \cos2\phi \end{pmatrix}$ & 2 & $E$ & $+$ \\
\air $\cos3\phi$ & 3 & $A_2$ & $-$ \\
\air $\sin3\phi$ & 3 & $A_1$ & $-$ \\
\moreair $\begin{pmatrix} \sin4\phi \\ -\cos4\phi \end{pmatrix}$ & 4 & $E$ & $+$ \\
\moreair $\begin{pmatrix} \cos5\phi \\ -\sin5\phi \end{pmatrix}$ & 5 & $E$ & $-$ \\
\air $\cos6\phi$ & 6 & $A_1$ & $+$ \\
\air $\sin6\phi$ & 6 & $A_2$ & $+$ \\ \hline\hline
\end{tabular}
\end{center}
\end{table}

We now discuss the irreps in turn. The superconducting state of full symmetry, which belongs to the trivial irrep $A_1$, was considered in \cite{VKF24}. The lowest-order basis function of $A_1$ is the constant function. It is also even under time reversal. Hence, there are no symmetry-im\-posed nodes. This is ``conventional'' \textit{s}-wave ($l=0$) pairing.
Recently, M\ae{}land \textit{et al.}\ \cite{MBT25} have proposed an $A_1$ pairing state with gap minima, but not nodes, at the arc centers, based on a phononic mechanism.

The lowest-order basis functions of $\E1$ and $\E2$ that are even under time reversal are $\sin2\phi$ and $\cos2\phi$, respectively. This can be described as \textit{d}-wave ($l=2$) pairing. Superpositions of the two components are also possible and preserve time-reversal symmetry if the coefficients of both can be made real simultaneously. All these functions break rotation symmetry: The pairing amplitude is not the same along all six Fermi arcs. The first component by itself as well as any symmetry-related state have point nodes with linear dispersion at some arc centers but not at all of them. There is no experimental indication for different gaps for different arcs~\cite{CSK24}.

There also exist natural states of $E$ symmetry that break time-reversal symmetry. They consist of a complex superposition of $\E1$ and $\E2$ components with a phase shift of $\pm \pi/2$ between them. These states can have isotropic gap magnitude but they do not have nodes. Moreover, there is no experimental indication for spontaneous breaking of time-reversal symmetry \cite{CSK24}.

The lowest-order basis function of $A_2$ that is even under time reversal is $\sin6\phi$. This is \textit{i}-wave ($l=6$) pairing \cite{CSK24} (the time-reversal odd function $\cos3\phi$ cannot appear, as discussed in Appendix \ref{app.symmetries}). It has symmetry-imposed point nodes with linear dispersion at all arc centers. In the absence of fine tuning, this is the only symmetry consistent with the ARPES experiments \cite{CSK24}. The nodes lead to topologically nontrivial Majorana cones and protected hinge states, as discussed in Ref.\ \cite{CSK24}. The gap function $\Delta(\phi)$ also changes sign between the arcs but these sign changes are not physically meaningful due to the absence of normal-state bands at the Fermi energy.

\begin{figure}
\raisebox{2.8cm}{(a)}\includegraphics[width=3.3cm]{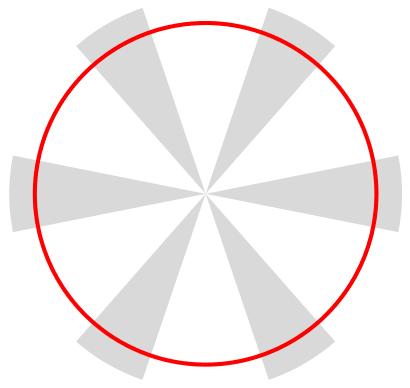}\hspace{-1.2em}$A_1$%
\raisebox{2.8cm}{(b)}\includegraphics[width=3.3cm]{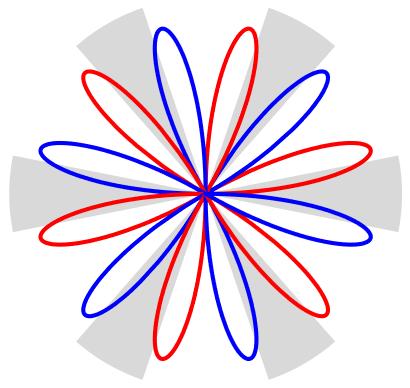}\hspace{-1.2em}$A_2$\\[1ex]
\raisebox{2.8cm}{(c)}\includegraphics[width=3.3cm]{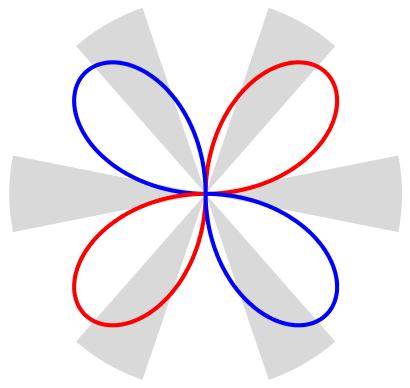}\hspace{-1.2em}$\E1$%
\raisebox{2.8cm}{(d)}\includegraphics[width=3.3cm]{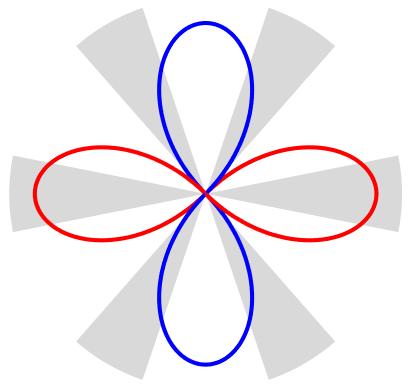}\hspace{-1.2em}$\E2$
\caption{\label{fig.basis}Polar plots of the lowest-order basis functions showing the symmetries of the superconducting pairing amplitudes of (a) $A_1$ symmetry ($\Delta(\phi) \propto 1$), (b) $A_2$ symmetry ($\sin6\phi$), (c) $\E1$ symmetry ($\sin2\phi$), and (d) $\E2$ symmetry ($\cos2\phi$). Red and blue color refer to positive and negative sign, respectively. The angular ranges spanned by the Fermi arcs are sketched as gray sectors, which have been exaggerated for clarity.}
\end{figure}

The lowest-order basis functions for all possible pairing symmetries are sketched in Fig.\ \ref{fig.basis}. The angular ranges of the Fermi arcs are also indicated.
Phase-sensitive measurements, like for the cuprates \cite{TsK00}, would be ideal to determine the symmetry of the superconducting OP. We are not aware of such measurements. In the following section, we propose that  in particular the $A_1$ and $A_2$ pairing states can be distinguished by their different reaction to applied magnetic fields.

\section{Ginzburg--Landau theory}
\label{sec.GL}

In this section, we set up the GL functional for the superconducting OPs. After discussing the principles used in constructing the functional, we give all terms up to fourth order in total and up to first order in magnetic field. Then, we employ this functional to predict effects of an applied magnetic field on $A_1$ and $A_2$ pairing.

\subsection{Principles of constructing the functional}
\label{sec.functional}

The pairing amplitude $\Delta(\vec k)$ in the projected BdG Hamiltonian is in general a superposition of basis functions of the irreps $A_1$, $A_2$ or irrep components $\E1$, $\E2$ with spatially varying complex coefficients. Their coarse-grained spatial structure is described by the GL functional. We refer to the coarse-grained coefficients as ``superconducting OPs'' and denote the OPs of $A_1$, $A_2$, and $E$ symmetry by $\psi_1$, $\psi_2$, and
\begin{equation}
\vec\psi_E = \begin{pmatrix}
  \psi_x \\ \psi_y
\end{pmatrix} ,
\end{equation}
respectively. They transform under the point group according to the irrep or irrep component and are mapped to the complex conjugate under time reversal.

The GL functional also contains the magnetic-field components, which transform like $R_x$, $R_y$, $R_z$ in Table~\ref{tab.char} under the point group. Thus, $(B_y,-B_x)$ transforms according to the irrep $E$ and $B_z$ according to $A_2$. Moreover, the magnetic field is odd under time reversal. Finally, we need the 2D gauge-covariant derivative
\begin{equation}
\vec D = \begin{pmatrix} D_x \\ D_y \end{pmatrix}
  = \vec\nabla - i\, \frac{q}{\hbar c}\, \vec A ,
\end{equation}
where $q = -2e$ is the charge of Cooper pairs and $\vec A$ is the in-plane part of the vector potential. $\vec D$ transforms according to the irrep $E$, as can easily be checked, i.e., it is an irreducible tensor operator of $E$. Since time reversal $\mathcal{T}$ does not affect the position but changes the sign of the magnetic field and the vector potential, $\mathcal{T}$ acts on the gauge-covariant derivative according to~\cite{endnote.TR}
\begin{equation}
\vec D = \vec\nabla - i\, \frac{q}{\hbar c}\, \vec A
  \stackrel{\mathcal{T}}{\longmapsto} \vec\nabla + i\, \frac{q}{\hbar c}\, \vec A = \vec D^\ast .
\label{3.TRS.D.2}
\end{equation}
We also require the derivative $\nabla_z = \partial/\partial z$ acting on the magnetic field. $\nabla_z$ transforms according to $A_1$ since $z$ does.

We first investigate which terms are possible in the GL functional. Terms in the functional must (a) respect the point-group symmetry, i.e., transform according to $A_1$, (b) respect time-reversal symmetry, (c) be real, and (d) be gauge invariant. The symmetry under global $\mathrm{U}(1)$ phase transformation is a special case of gauge invariance.

We construct all terms containing up to four factors of $\psi_1$, $\psi_2$, $\psi_x$, $\psi_y$, $\psi_1^\ast$, $\psi_2^\ast$, $\psi_x^\ast$, $\psi_y^\ast$, $B_x$, $B_y$, $B_z$, $D_x$, or $D_y$, with the following constraints: We only include terms that contain superconducting OPs or, in other words, we subtract the normal-state free energy. Due to gauge invariance, the number of complex conjugated OPs must equal the number of unconjugated OPs. We restrict ourselves to the first order in the magnetic field $\mathbf{B} = (B_x,B_y,B_z)$. This can be understood as the weak-field limit. However, some terms of higher order are generated by using the gauge-covariant derivative. We indicate two-component vectors by an arrow and three-com\-po\-nent vectors by bold-face font.

The strategy is the following: First, find all products of irreps or irrep components up to fourth order that result in $A_1$. Second, for each such product insert all combinations of $\psi_1$, $\psi_2$, $\psi_x$, $\psi_y$, $\psi_1^\ast$, $\psi_2^\ast$, $\psi_x^\ast$, $\psi_y^\ast$, $B_x$, $B_y$, $B_z$, $D_x$, and $D_y$ that transform like the factors, taking care of the constraints mentioned above. Third, generate real terms by taking the real or imaginary part. Some terms will turn out to vanish. Fourth, omit terms that are linearly dependent of other terms.

The first step can be done automatically, based on the possible expansions of irreps into second-order products,
\begin{align}
A_1 &= A_1 \otimes A_1 , \\
A_1 &= A_2 \otimes A_2 , \\
A_1 &= \E1 \otimes \E1 + \E2 \otimes \E2 , \\
A_2 &= A_1 \otimes A_2 , \\
A_2 &= A_2 \otimes A_1 , \\
A_2 &= \E1 \otimes \E2 - \E2 \otimes \E1 , \\
\begin{pmatrix} \E1 \\ \E2 \end{pmatrix}
  &= \begin{pmatrix} A_1 \otimes \E1 \\ A_1 \otimes \E2 \end{pmatrix} , \\
\begin{pmatrix} \E1 \\ \E2 \end{pmatrix}
  &= \begin{pmatrix} -A_2 \otimes \E2 \\ A_2 \otimes \E1 \end{pmatrix} , \\
\begin{pmatrix} \E1 \\ \E2 \end{pmatrix}
  &= \begin{pmatrix} \E1 \otimes A_1 \\ \E2 \otimes A_1 \end{pmatrix} , \\
\begin{pmatrix} \E1 \\ \E2 \end{pmatrix}
  &= \begin{pmatrix} -\E2 \otimes A_2 \\ \E1 \otimes A_2 \end{pmatrix} , \\
\begin{pmatrix} \E1 \\ \E2 \end{pmatrix}
  &= \begin{pmatrix} \E1 \otimes \E2 + \E2 \otimes \E1 \\
    \E1 \otimes \E1 - \E2 \otimes \E2 \end{pmatrix} .
\end{align}
By iterating these expansions, we obtain products of three or four factors. All products up to fourth order that transform according to $A_1$ are listed in Appendix \ref{app.products} for completeness.

Concerning the second step, we note that while the direct product of representations is commutative, the actual expressions involve the derivatives $D_x$, $D_y$, which do not commute with spatially-dependent fields. We here use the convention that the derivatives act on the next field to the right only.

For the third step, we examine how to construct terms that are real and even under time reversal. If only OPs and derivatives are present both conditions coincide since time reversal has the same effect as complex conjugation. Once one has found some term $X$ that respects the lattice symmetries and gauge invariance, one can easily make it time-reversal-even by adding its complex conjugate, $X + X^\ast$, essentially taking the real part. Note that if $X$ is imaginary the result will vanish.

For terms involving a magnetic-field component $B_j$, a different approach is necessary because the field is real but odd under time reversal. We need to make the part $X$ formed by derivatives and OPs odd under time reversal as well. This can be achieved by subtracting the complex conjugate. Since the resulting term is imaginary, we need to multiply by $i$, giving $i B_j\, (X - X^\ast)$, i.e., involving the imaginary part. This vanishes if $X$ is real.

The fourth step of eliminating redundant terms may also involve integrating by parts. For a generic term, integration by parts reads as
\begin{align}
\int &d^2r\, f^\ast(\vec r)\, D_j g(\vec r)
  = \int d^2r\, f^\ast \left( \nabla_j - i\, \frac{q}{\hbar c}\, A_j \right) g \nonumber \\
&= \int d^2r\, \left( - \nabla_j f^\ast - i\, \frac{q}{\hbar c}\, A_j f^\ast \right) g \nonumber \\
&= - \int d^2r\, g \left( \nabla_j + i\, \frac{q}{\hbar c}\, A_j \right) f^\ast
  = - \int d^2r\, g\, D_j^\ast f^\ast ,
\end{align}
where a surface term in the infinite has been dropped. Since we have already made all terms real we can apply complex conjugation without changing the result, giving
\begin{equation}
- \int d^2r\, g^\ast \left( \nabla_j - i\, \frac{q}{\hbar c}\, A_j \right) f
  = - \int d^2r\, g^\ast(\vec r)\, D_j f(\vec r) .
\end{equation}

\subsection{Ginzburg--Landau functional for PtBi\textsubscript{2}}

The GL functional is
\begin{equation}
\mathcal{F}[\psi_1,\psi_2,\vec\psi,\mathbf{B}]
  = \int d^2r\, F(\vec r) ,
\end{equation}
with
\begin{align}
&F = F_1 + F_2 + F_E + F_{12} + F_{1E} + F_{2E} + F_{12E} \nonumber \\
&\quad{}+ F^{(L)}_{1E} + F^{(L)}_{2E}
  + F^{(B)}_E + F^{(B)}_{12} + F^{(B)}_{1E} + F^{(B)}_{2E} \nonumber \\
&\quad{}+ F^{(j)}_E + F^{(j)}_{12} + F^{(j)}_{1E}
  + F^{(j)}_{2E} + F^{(j)\prime}_{1E}
  + F^{(j)\prime}_{2E} \nonumber \\
&\quad{}+ F^{(L)}_1 + F^{(L)}_2
  + F^{(L)}_E + F^{(L)}_{12} + F^{(L)\prime}_{1E} + F^{(L)\prime}_{2E} .
\end{align}
In the following, we construct these contributions to $F$.

The terms involving only one of the superconducting OPs and no explicit magnetic field (but the vector potential in $\vec D$) read as
\begin{align}
F_1 &= \frac{\alpha_1}{2}\, \psi_1^\ast \psi_1 + \frac{\beta_1}{4}\, (\psi_1^\ast \psi_1)^2
  + \frac{\gamma_1}{2}\, ( \vec D \psi_1 )^\ast \!\cdot ( \vec D \psi_1 ) ,
\label{def.F1} \\
F_2 &= \frac{\alpha_2}{2}\, \psi_2^\ast \psi_2 + \frac{\beta_2}{4}\, (\psi_2^\ast \psi_2)^2
  + \frac{\gamma_2}{2}\, ( \vec D \psi_2 )^\ast \!\cdot ( \vec D \psi_2 ) , \\
F_E &= \frac{\alpha_E}{2}\, \vec\psi_E^\ast \cdot \vec\psi_E
  + \frac{\beta_E}{4}\, (\vec\psi_E^\ast \cdot \vec\psi_E)^2 \nonumber \\
&\quad{}+ \frac{\beta_E'}{4}\, (\psi_x^\ast \psi_y - \psi_y^\ast \psi_x)^2
  + \frac{\gamma_E}{2}\, ( \vec D \cdot \vec\psi_E)^\ast ( \vec D \cdot \vec\psi_E )
  \nonumber \\
&\quad{}+ \frac{\gamma_E'}{2}\, ( D_x \psi_y - D_y \psi_x )^\ast ( D_x \psi_y - D_y \psi_x ) .
\label{def.FE}
\end{align}
The contributions $F_1$ and $F_2$ corresponding to the one-dimensional irreps $A_1$ and $A_2$ have the standard form. $F_E$ contains more terms since both $\vec\psi_E$ and $\vec D$ are $E$ doublets and there are three ways to construct a product of full symmetry out of four $E$ factors:
\begin{align}
&(\E1 \otimes \E1 + \E2 \otimes \E2) \otimes (\E1 \otimes \E1 + \E2 \otimes \E2) , \\
&(\E1 \otimes \E2 - \E2 \otimes \E1) \otimes (\E1 \otimes \E2 - \E2 \otimes \E1) , \\
&(\E1 \otimes \E1 - \E2 \otimes \E2) \otimes (\E1 \otimes \E1 - \E2 \otimes \E2) \nonumber \\
&\quad{}+ (\E1 \otimes \E2 + \E2 \otimes \E1) \otimes (\E1 \otimes \E2 + \E2 \otimes \E1) ,
\end{align}
which result from expanding $A_1 \otimes A_1$, $A_2 \otimes A_2$, and $E \otimes E$, respectively. The corresponding terms of fourth order in $\vec\psi_E$ are proportional to
\begin{align}
(\vec\psi_E^\ast &\cdot \vec\psi_E)^2 , \\
(\psi_x^\ast \psi_y &- \psi_y^\ast \psi_x)^2 , \\
(\psi_x^\ast \psi_x - \psi_y^\ast \psi_y)^2 &+ (\psi_x^\ast \psi_y + \psi_y^\ast \psi_x)^2 ,
\end{align}
respectively. However, these expressions are not linearly independent since the last expression equals
\begin{equation}
(\psi_x^\ast \psi_x + \psi_y^\ast \psi_y)^2
  + (\psi_x^\ast \psi_y - \psi_y^\ast \psi_x )^2 ,
\end{equation}
which is the sum of the other two. This reduction is due to the invariance under interchange of the two $\psi_j^\ast$ or of the two $\psi_j$ factors. Moreover, we do not obtain additional products by using the ordering $\psi^\ast \psi^\ast \psi \psi$.

For the terms of second order in $\vec\psi_E$ and containing two derivatives, the situation is more complicated since the gauge-covariant derivatives generally do not commute. The allowed terms corresponding to the products $A_1 \otimes A_1$, $A_2 \otimes A_2$, and $E \otimes E$ can be written as
\begin{align}
(\vec D\cdot \vec\psi_E)^\ast &(\vec D\cdot \vec\psi_E) ,
\label{3.DpDp.3a} \\
(D_x \psi_y - D_y \psi_x)^\ast &(D_x \psi_y - D_y \psi_x) ,
\label{3.DpDp.3b} \\
(D_x \psi_x - D_y \psi_y)^\ast &(D_x \psi_x - D_y \psi_y) \nonumber \\
{}- (D_x \psi_y + D_y &\psi_x)^\ast (D_x \psi_y + D_y \psi_x) ,
\label{3.DpDp.3c}
\end{align}
respectively. They all satisfy time-reversal symmetry, see Eq.\ (\ref{3.TRS.D.2}). For a term $(D_i f)^\ast D_k g$, an elementary calculation involving twofold integration by part gives
\begin{align}
\int &d^2r\, (D_i f)^\ast\, D_j g \nonumber \\
&= \int d^2r \left[
    (D_j f)^\ast\, D_i g + i\,\frac{q}{\hbar c}\, (\nabla_i A_j - \nabla_j A_i)\, f^\ast g \right]
    \nonumber \\
&= \int d^2r \left[
    (D_j f)^\ast\, D_i g + i\,\frac{q}{\hbar c}\, \sum_k \epsilon_{ijk}\, B_k\, f^\ast g \right] ,
\end{align}
where $\epsilon_{ijk}$ is the Levi--Civita symbol. The application to Eq.\ (\ref{3.DpDp.3c}) generates, under the spatial integral, the terms in Eqs.\ (\ref{3.DpDp.3a}) and (\ref{3.DpDp.3b}) plus the term
\begin{align}
- 4i\,&\frac{q}{\hbar c} \left( \nabla_x A_y - \nabla_y A_x \right)
  (\psi_x^\ast \psi_y - \psi_y^\ast \psi_x) \nonumber \\
&= - 4i\, \frac{q}{\hbar c}\, B_z\, (\psi_x^\ast \psi_y - \psi_y^\ast \psi_x) ,
\end{align}
which appears if a magnetic field is applied along the \textit{z} direction. It is included below in Eq.~(\ref{eq.FB.E}).

The coupling terms without explicit magnetic field and of even order in derivatives read as
\begin{align}
F_{12} &= \frac{\delta_{12}}{2}\, \psi_1^\ast \psi_1 \psi_2^\ast \psi_2
  + \frac{\delta'_{12}}{2} \left( \psi_1^\ast \psi_2 \psi_1^\ast \psi_2 + \mathrm{c.c.} \right) ,
\label{def.F12} \\
F_{1E} &= \frac{\delta_{1E}}{2}\, \psi_1^\ast \psi_1 \vec\psi_E^\ast \cdot \vec\psi_E
  + \frac{\delta'_{1E}}{2}\, \big( \psi_1^\ast \vec\psi_E \cdot \psi_1^\ast \vec\psi_E
  + \mathrm{c.c.} \big) \nonumber \\
&\quad{}+ \frac{\delta''_{1E}}{2}\, \big[ \psi_1^\ast \psi_x ( \psi_x^\ast \psi_y
    + \psi_y^\ast \psi_x ) \nonumber \\
&\qquad{}+ \psi_1^\ast \psi_y ( \psi_x^\ast \psi_x
    - \psi_y^\ast \psi_y ) + \mathrm{c.c.} \big] \nonumber \\
&\quad{}+ \frac{\epsilon_{1E}}{2}\, \big[ ( D_x \psi_1 )^\ast ( D_x \psi_y + D_y \psi_x ) \nonumber \\
&\qquad{}+ ( D_y \psi_1 )^\ast ( D_x \psi_x - D_y \psi_y ) + \mathrm{c.c.} \big] , \\
F_{2E} &= \frac{\delta_{2E}}{2}\, \psi_2^\ast \psi_2 \vec\psi_E^\ast \cdot \vec\psi_E
  + \frac{\delta'_{2E}}{2}\, \big( \psi_2^\ast \vec\psi_E \cdot \psi_2^\ast \vec\psi_E
  + \mathrm{c.c.} \big) \nonumber \\
&\quad{}+ \frac{\delta''_{2E}}{2}\, \big[ \psi_2^\ast \psi_x ( \psi_x^\ast \psi_x
  - \psi_y^\ast \psi_y ) \nonumber \\
&\qquad{}- \psi_2^\ast \psi_y ( \psi_x^\ast \psi_y
  + \psi_y^\ast \psi_x ) + \mathrm{c.c.} \big] \nonumber \\
&\quad{}+ \frac{\epsilon_{2E}}{2}\, \big[ ( D_x \psi_2 )^\ast ( D_x \psi_x - D_y \psi_y ) \nonumber \\
&\qquad{}- ( D_y \psi_2 )^\ast ( D_x \psi_y + D_y \psi_x ) + \mathrm{c.c.} \big] , \\
F_{12E} &= \frac{\delta_{12E}}{2} \left[ \psi_1^\ast \psi_2\,
    (\psi_x^\ast \psi_y - \psi_y^\ast \psi_x) + \mathrm{c.c.} \right] .
\end{align}
A term proportional to $( D_x \psi_1 )^\ast D_y \psi_2 - ( D_y \psi_1 )^\ast D_x \psi_2 + \mathrm{c.c.}$ is also symmetry allowed. In analogy to the terms in $F_E$ with two derivatives, twofold integration by parts gives, under the spatial integral, $i\, ({q}/{\hbar c})\, B_z \left( \psi_1^\ast \psi_2 - \psi_2^\ast \psi_1 \right)$, which is included below in Eq.~(\ref{DF12.3}). Moreover, the sym\-me\-try-al\-lowed term proportional to $\psi_1^\ast \psi_x \psi_2^\ast \psi_y - \psi_1^\ast \psi_y \psi_2^\ast \psi_x + \mathrm{c.c.}$ vanishes.

Terms of first order in derivatives are called ``Lifshitz invariants'' \cite{MiS94, Ede96, Agt03, MiS08, SSY17}. We find the following Lifshitz invariants without explicit magnetic field:
\begin{align}
F^{(L)}_{1E} &= \eta_{1E} \left( \psi_1^\ast \vec D \cdot \vec \psi_E
  - \vec \psi_E^\ast \cdot \vec D \psi_1 \right) , \\
F^{(L)}_{2E} &= \eta_{2E} \left[ \psi_2^\ast ( D_x \psi_y - D_y \psi_x )
- ( \psi_y^\ast D_x - \psi_x^\ast D_y ) \psi_2 \right] .
\end{align}
They are enabled by the possibility of products of OPs that transform in the same way as the derivative. Lifshitz invariants allow for unconventional superconducting states, as discussed below.

In the presence of a magnetic field, we obtain the following terms of first order in the field and without any derivatives:
\begin{align}
F^{(B)}_E &= \zeta_E\, i\, B_z\, ( \psi_x^\ast \psi_y - \psi_y^\ast \psi_x ) ,
\label{eq.FB.E} \\
F^{(B)}_{12} &= \zeta_{12}\, i\, B_z\, ( \psi_1^\ast \psi_2 - \psi_2^\ast \psi_1 ) ,
\label{DF12.3} \\
F^{(B)}_{1E} &= \zeta_{1E}\, i\, \big[ B_y\, ( \psi_1^\ast \psi_x - \psi_x^\ast \psi_1 )
  \nonumber \\
&\quad{}- B_x\, ( \psi_1^\ast \psi_y - \psi_y^\ast \psi_1 ) \big] , \\
F^{(B)}_{2E} &= \zeta_{2E}\, i\, \big[ B_y\, ( \psi_2^\ast \psi_y - \psi_y^\ast \psi_2 )
  \nonumber \\
&\quad{}+ B_x\, ( \psi_2^\ast \psi_x - \psi_x^\ast \psi_2 ) \big] .
\label{eq.FB.2E}
\end{align}
We have seen above that terms of the form of $F^{(B)}_E$ and $F^{(B)}_{12}$ are generated from terms with two gauge-covariant derivatives. However, terms containing the magnetic field generically stem from two distinct physical effects: One is the orbital motion of the charged fields $\psi_1$, $\psi_2$, and $\vec\psi_E$, expressed by the gauge-covariant derivatives. The other is the spin of the Cooper pairs. Since PtBi\textsubscript{2} lacks inversion symmetry even in the bulk and in particular at the surface we expect all superconducting states to exhibit singlet-triplet mixing \cite{BaS12, SSY17}, i.e., nonvanishing Cooper-pair spin.

Moreover, there are contributions involving derivatives of the magnetic field, which, according to the Amp\`ere--Maxwell law, are related to currents. To fourth order overall, these contributions can only be of the form $i\,(\nabla B)(\psi^\ast\psi - \mathrm{c.c.})$ in order to respect time-reversal symmetry. Several allowed terms vanish because the part $\psi^\ast\psi$ is real. The remaining contributions involving in-plane derivatives are
\begin{align}
F^{(j)}_E &= \kappa_E\, i\, (\nabla_x B_x + \nabla_y B_y) (\psi_x^\ast \psi_y
  - \psi_y^\ast \psi_x) ,
\label{eq.FBp.E} \\
F^{(j)}_{12} &= \kappa_{12}\, i\, (\nabla_x B_x + \nabla_y B_y) (\psi_1^\ast \psi_2
  - \psi_2^\ast \psi_1) ,
\label{eq.FBp.12} \\
F^{(j)}_{1E} &= \kappa_{1E}\, i\, \big[ (\nabla_y B_z)
    (\psi_1^\ast \psi_x - \psi_x^\ast \psi_1) \nonumber \\
&\qquad{}- (\nabla_x B_z) (\psi_1^\ast \psi_y - \psi_y^\ast \psi_1) \big] \nonumber \\
&\quad{}+ \kappa_{1E}'\, i\, \big[ (-\nabla_x B_x + \nabla_y B_y)
    (\psi_1^\ast \psi_x - \psi_x^\ast \psi_1) \nonumber \\
&\qquad{}+ (\nabla_x B_y + \nabla_y B_x) (\psi_1^\ast \psi_y - \psi_y^\ast \psi_1) \big] , \\
F^{(j)}_{2E} &= \kappa_{2E}\, i \big[ (\nabla_y B_z)
    (\psi_2^\ast \psi_y - \psi_y^\ast \psi_2) \nonumber \\
&\qquad{}+ (\nabla_x B_z) (\psi_2^\ast \psi_x - \psi_x^\ast \psi_2) \big] \nonumber \\
&\quad{}+ \kappa_{2E}'\, i\, \big[ (\nabla_x B_x - \nabla_y B_y)
    (\psi_2^\ast \psi_y - \psi_y^\ast \psi_2) \nonumber \\
&\qquad{}+ (\nabla_x B_y + \nabla_y B_x) (\psi_2^\ast \psi_x - \psi_x^\ast \psi_2) \big] .
\end{align}
Unlike the OPs of the 2D superconductor, the magnetic field $\mathbf{B}$ can have a nonzero derivative in the \textit{z} direction. Since $\nabla_z$ is invariant under the point group ($A_1$ symmetry) the resulting contributions can simply be obtained from Eqs.\ (\ref{eq.FB.E})--(\ref{eq.FB.2E}) by replacing $B_j$ by $\nabla_z B_j$. Due to Gauss' law for the magnetic field, the results involving $\nabla_z B_z$ reproduce $F^{(j)}_E$ and $F^{(j)}_{12}$. The new terms are
\begin{align}
F^{(j)\prime}_{1E} &= \lambda_{1E}\, i\, \big[
  (\nabla_z B_y) ( \psi_1^\ast \psi_x - \psi_x^\ast \psi_1 ) \nonumber \\
&\quad{}- (\nabla_z B_x) ( \psi_1^\ast \psi_y - \psi_y^\ast \psi_1 ) \big] , \\
F^{(j)\prime}_{2E} &= \lambda_{2E}\, i\, \big[
  (\nabla_z B_y) ( \psi_2^\ast \psi_y - \psi_y^\ast \psi_2 ) \nonumber \\
&\quad{}+ (\nabla_z B_x) ( \psi_2^\ast \psi_x - \psi_x^\ast \psi_2 ) \big] .
\end{align}
Since the superconductor exists at the surface the derivative $\nabla_z \mathbf{B}$ is generically nonzero if a field is applied.

Finally, we find contributions of first order in the field and in the derivative acting on OPs, hence, additional Lifshitz invariants \cite{MiS94, Ede96, Agt03, MiS08, SSY17}. These terms read as
\begin{align}
F^{(L)}_1 &= \xi_1\, i \left( B_y \psi_1^\ast D_x \psi_1 - B_x \psi_1^\ast D_y \psi_1
  - \mathrm{c.c.} \right) , \\
F^{(L)}_2 &= \xi_2\, i \left( B_y \psi_2^\ast D_x \psi_2 - B_x \psi_2^\ast D_y \psi_2
  - \mathrm{c.c.} \right) , \\
F^{(L)}_E &= \xi_E\, i B_z \big[ \psi_x^\ast ( D_x \psi_x - D_y \psi_y )
  - \psi_y^\ast ( D_x \psi_y + D_y \psi_x ) \big] \nonumber \\
&\quad{}+ \xi_E'\, i \left[ ( B_y \psi_x^\ast - B_x \psi_y^\ast ) ( D_x \psi_x + D_y \psi_y )
  - \mathrm{c.c.} \right] \nonumber \\
&\quad{}+ \xi_E''\, i \left[ ( B_y \psi_y^\ast + B_x \psi_x^\ast ) ( D_x \psi_y - D_y \psi_x )
  - \mathrm{c.c.} \right] ,
\label{6.DFEprime} \\
F^{(L)}_{12} &= \xi_{12}\, i \left( B_x \psi_1^\ast D_x \psi_2
  + B_y \psi_1^\ast D_y \psi_2 - \mathrm{c.c.} \right) , \\
F^{(L)\prime}_{1E} &= \xi_{1E}\, i \left[ B_z \psi_1^\ast ( D_x \psi_y - D_y \psi_x ) - \mathrm{c.c.}
    \right] \nonumber \\
&\quad{}+ \xi_{1E}'\, i\, \big[ B_y \psi_1^\ast ( D_x \psi_y + D_y \psi_x ) \nonumber \\
&\qquad{}- B_x \psi_1^\ast ( D_x \psi_x - D_y \psi_y ) - \mathrm{c.c.} \big] , \\
F^{(L)\prime}_{2E} &= \xi_{2E}\, i \left[ B_z \psi_2^\ast ( D_x \psi_x + D_y \psi_y ) - \mathrm{c.c.}
  \right] \nonumber \\
&\quad{}+ \xi_{2E}'\, i\, \big[ B_y \psi_2^\ast ( D_x \psi_x - D_y \psi_y ) \nonumber \\
&\qquad{}+ B_x \psi_2^\ast ( D_x \psi_y + D_y \psi_x ) - \mathrm{c.c.} \big] .
\end{align}
The two terms in $F^{(L)}_E$ with in-plane field components are similar to the terms with two derivatives in $F_E$, see Eq.\ (\ref{def.FE}). The terms corresponding to the products $A_1 \otimes A_1$, $A_2 \otimes A_2$, and $E \otimes E$ read as
\begin{align}
&i\, \big[ ( B_y \psi_x^\ast - B_x \psi_y^\ast ) ( D_x \psi_x + D_y \psi_y )
  - \mathrm{c.c.} \big] ,
\label{disc.A1A1.3} \\
&i\, \big[ ( B_y \psi_y^\ast + B_x \psi_x^\ast ) ( D_x \psi_y - D_y \psi_x )
  - \mathrm{c.c.} \big] , \\
&i\, \big[ ( B_y \psi_x^\ast + B_x \psi_y^\ast ) ( D_x \psi_x - D_y \psi_y ) \nonumber \\
&\quad{}+ ( B_y \psi_y^\ast - B_x \psi_x^\ast ) ( D_x \psi_y + D_y \psi_x ) - \mathrm{c.c.} \big] ,
\label{disc.EE.3}
\end{align}
respectively. One can show that taking the sum of the first two terms and using integration by parts generates the third term plus a field-derivative term proportional to $F^{(j)}_E$ in Eq.\ (\ref{eq.FBp.E})~\cite{endnote.FLE}.

\subsection{$A_1$ and $A_2$ pairing in a magnetic field}

As noted above, high-precision ARPES results \cite{CSK24} are most consistent with $A_2$ pairing in PtBi\textsubscript{2} but cannot exclude $A_1$ pairing with accidental (near) nodes and fine-tuned linear dispersion. It is thus important to find experimental signatures that distinguish between these possibilities. Therefore, we focus on the comparison of primary $A_1$ and $A_2$ superconducting OPs.

As a reminder, in the absence of a magnetic field, $A_1$ pairing is stabilized if the coefficient $\alpha_1$ in $F_1$, Eq.\ (\ref{def.F1}), becomes negative below a certain critical temperature $T_c$, while $\alpha_2$ is still positive. The coefficients $\beta_1$, $\gamma_1$, $\beta_2$, and $\gamma_2$ must be positive to ensure that the GL functional is bounded from below. Even if $\alpha_2$ also becomes negative at a lower temperature, $A_2$ pairing generically does not set in immediately because the coupling between $A_1$ and $A_2$ order in $F_{12}$, Eq.\ (\ref{def.F12}), is expected to be repulsive because the superconducting states of different symmetry compete for the same electrons. The case of primary $A_2$ pairing is analogous.

First, we consider a uniform magnetic field applied in parallel to the surface. A uniform field is indeed expected in this case if the two surfaces are decoupled. For $A_2$ superconductivity and a magnetic field along the \textit{x} direction, i.e., parallel to the $\Gamma$M lines and to the nodal planes, the relevant terms containing the magnetic field in the GL functional are
\begin{align}
F^{(B)}_{2E} &= \zeta_{2E}\, i\, B_x\, (\psi_2^\ast \psi_x - \psi_x^\ast \psi_2) , \\
F^{(L)}_2 &= -\xi_2\, i\, B_x\, (\psi_2^\ast \nabla_y \psi_2 - \psi_2 \nabla_y \psi_2^\ast) .
\end{align}
We have replaced $\vec D$ by $\vec\nabla$ since we treat the magnetic field at first order. The second term is discussed below in the context of helical states. The first one implies that a superconducting OP $\psi_x$ is induced, which, for weak fields, is linear in $B_x$ and in $\psi_2$. Moreover, since $F^{(B)}_{2E}$ contains the imaginary part of $\psi_2^\ast \psi_x$ the induced OP $\psi_x$ has a phase shift of $\pm \pi/2$ relative to the primary OP $\psi_2$. The sign depends on the sign of $\zeta_{2E} B_x$. The induced first component of the $E$ OP has the symmetry of $\sin2\phi$, see Table \ref{tab.fphi}. Hence, it has nodes at the Fermi-arc centers at $\phi=0$ and $\phi=\pi$ but is nonzero on the other four arcs. It thus preserves the nodes at $\phi=0$, $\pi$ and removes the nodes at $\phi=\pi/3$, $2\pi/3$, $4\pi/3$, $5\pi/3$~\cite{endnote.energygap}.

On the other hand, for $A_1$ superconductivity and field along the \textit{x} direction, the relevant terms are
\begin{align}
F^{(B)}_{1E} &= -\zeta_{1E}\, i\, B_x\, (\psi_1^\ast \psi_y - \psi_y^\ast \psi_1) , \\
F^{(L)}_1 &= -\xi_1\, i\, B_x\, (\psi_1^\ast \nabla_y \psi_1 - \psi_1 \nabla_y \psi_1^\ast) .
\end{align}
The OP induced by $F^{(B)}_{1E}$ is $\psi_y$, which has the symmetry of $\cos2\phi$. This OP does not have nodes at any arc center. If the primary $A_1$ superconductivity has accidental nodes, as required by the ARPES results \cite{CSK24}, they are all removed by the applied field. Hence, we find qualitatively different behavior for $A_2$ compared to $A_1$ superconductivity. It is easy to check that for magnetic field along the \textit{y} direction, i.e., between the Fermi arcs, the results are reversed.

For magnetic field applied in the \textit{z} direction, the situation is more complicated because of the formation of vortices. Our GL functional in principle allows to obtain vortex solutions. Due to the presence of nonzero derivatives, many of the terms are activated. Here, we only discuss the case of approximately uniform field close to but below the upper critical field, when the magnetic flux associated with individual vortices strongly overlaps. In this case, the relevant term for both $A_1$ and $A_2$ superconductivity is
\begin{equation}
F^{(B)}_{12} = \zeta_{12}\, i\, B_z\, (\psi_1^\ast \psi_2 - \psi_2^\ast \psi_1) .
\end{equation}
Hence, for both cases the other OP is induced with a phase difference of $\pm \pi/2$. For primary $A_2$ OP, the induced $A_1$ OP is the same at all arcs and generically nodeless, and thus removes all nodes. For primary $A_1$ OP with accidental nodes, the induced $A_2$ OP has symmetry-imposed nodes at all arc centers. Hence, all accidental nodes persist. Again, we find qualitatively different behavior for $A_1$ and $A_2$ pairing.

The ideal experiment to probe the persistence or lifting of the nodes would be ARPES in applied magnetic field. This is difficult because the magnetic field affects the photoelectrons' trajectory. However, Ryu \textit{et al.}\ \cite{RRJ23} have recently made significant progress by confining the magnetic field to a thin layer close to the sample surface.

The most promising idea, in particular if all nodes are removed, may be to observe a change of the shape of the tunneling gap in STS upon removal of the nodes by an applied magnetic field. In fact, the results for zero magnetic field are consistent with a nodal gap~\cite{SFH24}.

Next, we address Lifshitz invariants activated by the applied magnetic field. Besides the terms $F^{(L)}_1$ and $F^{(L)}_2$, additional terms become nonzero because of the presence of the induced superconducting OP. However, the induced OP is of first order in magnetic field so that terms that contain explicit factors of the field or more than one factor of the induced OP are of higher order in the field and should be ignored for consistency. A magnetic field along the \textit{z} direction does not activate Lifshitz invariants for primary $A_1$ or $A_2$ pairing, to linear order in the field. For a magnetic field along the \textit{x} direction, the remaining terms are the Lifshitz invariants $F^{(L)}_{1E}$ and $F^{(L)}_{2E}$. We obtain, for $A_2$ pairing,
\begin{align}
F^{(L)}_2 + F^{(L)}_{2E} &=
  - \xi_2\, i\, B_x\, (\psi_2^\ast \nabla_y \psi_2 - \psi_2 \nabla_y \psi_2^\ast) \nonumber \\
&\quad{}- \eta_{2E} \left( \psi_2^\ast \nabla_y \psi_x - \psi_x^\ast \nabla_y \psi_2 \right) .
\end{align}
The ansatz $\psi_2 = |\psi_2|\, e^{i q_y y}$, $\psi_x = \pm i\, |\psi_x|\, e^{i q_y y}$ gives
\begin{equation}
F^{(L)}_2 + F^{(L)}_{2E} = 2\xi_2 B_x\, q_y\, |\psi_2|^2
  \pm 2 \eta_{2E}\, q_y\, |\psi_2||\psi_x| .
\label{eq.helical.3}
\end{equation}
In addition, the free energy contains a standard term proportional to $q_y^2$. Minimization with respect to $q_y$ leads to a helical state \cite{ADE20} with modulation vector $\vec q = (0,q_y)$ proportional to $\xi_2 B_x\, |\psi_2|^2 \pm \eta_{2E}\, |\psi_2||\psi_x|$ and thus to the applied field, as expected for Lifshitz invariants \cite{MiS94, Ede96, Agt03, MiS08, SSY17}. For primary $A_1$ pairing, we analogously have to consider
\begin{align}
F^{(L)}_1 + F^{(L)}_{1E} &=
  - \xi_1\, i\, B_x\, (\psi_1^\ast \nabla_y \psi_1 - \psi_1 \nabla_y \psi_1^\ast) \nonumber \\
&\quad{}+ \eta_{1E} \left( \psi_1^\ast \nabla_y \psi_y - \psi_y^\ast \nabla_y \psi_1 \right)
  \nonumber \\
&= 2\xi_1 B_x\, q_y\, |\psi_1|^2 \mp 2 \eta_{1E}\, q_y\, |\psi_1||\psi_y| .
\label{eq.helical.4}
\end{align}
Minimization leads to a helical state with modulation vector proportional to $\xi_1 B_x\, |\psi_1|^2 \mp \eta_{1E}\, |\psi_1||\psi_y|$. Equations (\ref{eq.helical.3}) and (\ref{eq.helical.4}) show that the two solutions for signs $\pm$ in the ansatz have different free energies. Which one is stable depends on the signs of the coefficients and on the direction of the magnetic field. Hence, we obtain a modulation with a single wave vector $\vec q$, i.e., a helical state \cite{SSY17}. For both $A_2$ and $A_1$ symmetry, the modulation vector is proportional to the magnetic-field strength and perpendicular to the field so that it does not help to distinguish $A_2$ from $A_1$ pairing. For convenience, we summarize the predictions for $A_1$ vs.\ $A_2$ pairing in Table~\ref{tab.summary}.

\begin{table}
\begin{center}
\caption{\label{tab.summary}Summary of the consequences of a uniform magnetic field applied along the \textit{x} vs.\ \textit{z} direction for pairing with $A_1$ symmetry and with $A_2$ symmetry. Nodes for $A_1$ symmetry are accidental, while they are required by $A_2$ symmetry.}
\begin{tabular}{c@{\quad}c@{\qquad}c} \hline\hline
\air Field & $A_1$ pairing & $A_2$ pairing \\ \hline
& \air all nodes gapped out & two nodes gapped out \\ \cline{2-3}
\raisebox{1.5ex}[-1.5ex]{$B_x$} &
\multicolumn{2}{c}{\air helical state, modulation vector $(0,q_y) \propto B_x$} \\ \hline
& \air all nodes persist &  all nodes gapped out \\ \cline{2-3}
\raisebox{1.5ex}[-1.5ex]{$B_z$} & \multicolumn{2}{c}{\air no modulation} \\ \hline\hline
\end{tabular}
\end{center}
\end{table}

\section{Summary and conclusions}
\label{sec.summary}

In this work, we have constructed the GL functional for trigonal PtBi\textsubscript{2} including superconducting OPs of all possible point-group symmetries (irreps $A_1$, $A_2$, and $E$ of point group $C_{3v}$) in the presence of a magnetic field. Many unconventional terms are generated, including Lifshitz invariants, i.e., terms of first order in gradients of OPs.

As an application, we have studied the behavior of the nodes in an applied uniform magnetic field. We find that the results are distinct for $A_1$ pairing with accidental nodes and $A_2$ pairing with symmetry-imposed nodes. For a uniform field applied along the \textit{x} direction, i.e., the direction towards the Fermi-arc centers, $A_2$ pairing is characterized by the nodes in the direction parallel to the field being preserved whereas the other four nodes are gapped out. On the other hand, for $A_1$ pairing all nodes are gapped out. For an approximately uniform field applied along the \textit{z} direction, $A_2$ pairing is characterized by all nodes being gapped out, whereas for $A_1$ pairing all nodes persist.

These predictions could, in principle, be tested by magnetoARPES \cite{RRJ23}, which is technically challenging. It should also be possible to distinguish between a full and a nodal gap based on measurements of the tunneling gap in STS. An in-plane field also generates spatially modulated superconductivity with a single modulation vector $\vec q$, i.e., a helical state, for both $A_1$ and $A_2$ pairing.

The inclusion of gradients also allows to describe both vortex states and local effects of impurities. We expect that the activation of many terms in the GL functional leads to a complex admixture of superconducting OPs. For example, the Lifshitz invariants containing derivatives but no fields, $F^{(L)}_{1E}$ and $F^{(L)}_{2E}$, induce a large $E$ OP in the vicinity of vortex cores, where the primary $A_1$ or $A_2$ OP changes rapidly. We leave the study of vortices in PtBi\textsubscript{2} to future work. It may shed light on the question why vortices have so far not been observed in scanning SQUID experiments~\cite{SFH24}.

Another issue for future study is the interplay between bulk and surface superconductivity in trigonal PtBi\textsubscript{2}. As noted above, bulk superconductivity sets in at a critical temperature on the order of $1\,\mathrm{K}$ \cite{SKP20, ZCC24}. The bulk OP could similarly be decomposed into $A_1$, $A_2$, and $E$ components and the coupling between bulk and surfaces could be described within the Ginzburg--Landau framework. We expect that bulk superconductivity strongly affects surface superconductivity. For example, if bulk superconductivity were of (conventional) $A_1$ symmetry, whereas surface superconductivity transforms according to $A_2$, the bulk OP would open an $A_1$ gap on the Fermi arcs by proximity effect, which would lower the possible free-energy gain for surface superconductivity and thereby suppress it.

Of course, while much can be learned from symmetry-based phenomenological GL theory, one would also like to understand the microscopic mechanism of superconductivity in PtBi\textsubscript{2}. This requires microscopic modeling. An interesting aspect is the connection to bulk states. The localization length of the Fermi-arc states diverges towards the projections of the Weyl points. Experimentally, the pairing amplitude appears to vanish in this limit \cite{CSK24}. On the other hand, the band splitting also goes to zero so that multiband pairing becomes possible.

To conclude, the results from the symmetry-based phenomenological GL theory presented here can help to elucidate various scenarios for the surface-superconducting state of trigonal PtBi\textsubscript{2}, including nodal topological \textit{i}-wave superconductivity. However, many fundamental and microscopic open questions remain, in particular regarding the mechanism that drives surface superconductivity in PtBi\textsubscript{2}.

\acknowledgments

The authors thank Sergey Borisenko, Ion Cosma Fulga, Christian Hess, Julia M. Link, and Jos\'e Lorenzana for useful discussions. Financial support by Deutsche Forschungsgemeinschaft, in part through Collaborative Research Center SFB 1143, project A04, project id 247310070, and W\"urzburg-Dresden Cluster of Excellence ct.qmat, EXC 2147, project id 390858490, is gratefully acknowledged.

\appendix

\section{Symmetries of pairing states}
\label{app.symmetries}

In this appendix, we give more details on the derivation of possible symmetries of the superconducting gap function. In the BdG Hamiltonian $\mathcal{H}(\vec k)$ in Eq.\ (\ref{eq.BdG.2}), the off-diagonal block $\hat\Delta(\vec k)$ is a $2\times 2$ matrix on spin space. It is useful to write $\hat\Delta(\vec k) = \hat D(\vec k) \hat U_T$, where $\hat U_T$ is the unitary part of the anti-unitary time-reversal operator, here $\hat U_T = i\sigma_y$. One can show that $\hat D(\vec k)$ transforms like a matrix under point-group transformations \cite{TiB21}, which allows one to decompose $\hat D(\vec k)$ into contributions transforming according to the irreps $A_1$, $A_2$, and $E$ (i.e., being irreducible tensor operators belonging to these irreps).

As stated in Sec.\ \ref{sec.nodal}, fermionic antisymmetry additionally requires~\cite{Gor58, SiU91, CTS16, TiB21}
\begin{equation}
\hat\Delta^T(-\vec k) = -\hat\Delta(\vec k) .
\label{eq.A.1aa}
\end{equation}
For the matrix $\hat D(\vec k)$, this condition translates to
\begin{align}
\hat U_T \hat D^T(-\vec k) \hat U_T^\dagger
&= \hat U_T \hat U_T^* \hat\Delta^T(-\vec k) \hat U_T^\dagger
  = - \hat\Delta^T(-\vec k) \hat U_T^\dagger \nonumber \\
&= + \hat\Delta(\vec k) \hat U_T^\dagger = \hat D(\vec k) .
\label{eq.A.1a}
\end{align}
Note that $\hat D^T$ generally differs from $\hat D^*$ since $\hat D$ is not Hermitian. For a Hermitian matrix, Eq.\ (\ref{eq.A.1a}) would just express time-reversal symmetry. For the generally non-Hermitian $\hat D$, one gets distinct relations containing $\hat D^T$ and $\hat D^*$. The derivation gives the form with $\hat D^T$.

As a $2\times 2$ matrix, $\hat D(\vec k)$ can also be written as a linear combination of the identity matrix $\sigma_0$ and the Pauli matrices $\mbox{\boldmath$\sigma$} = (\sigma_x, \sigma_y, \sigma_z)$,
\begin{equation}
\hat D(\vec k) = \psi(\vec k)\, \sigma_0 + \mathbf{d}(\vec k) \cdot \mbox{\boldmath$\sigma$} .
\end{equation}
Here, $\psi$ is the pairing amplitude for spin-singlet pairing, while $\mathbf{d}$ describes the strength and spin content of spin-triplet pairing. $\sigma_0$ is even under time reversal, whereas $\mbox{\boldmath$\sigma$}$ is odd. The condition (\ref{eq.A.1a}) thus implies that
\begin{align}
\psi(-\vec k) &= \psi(\vec k) ,
\label{eq.A.1b1} \\
\mathbf{d}(-\vec k) &= -\mathbf{d}(\vec k) ,
\label{eq.A.1b2}
\end{align}
as is well known.

As noted in Table \ref{tab.char}, the spin angular-momentum component $\sigma_z$ transforms according to $A_2$, while $(\sigma_y,-\sigma_x)$ transforms as an $E$ doublet. $\sigma_0$ of course transforms trivially according to $A_1$. Choosing functions $\psi(\vec k)$ and $\mathbf{d}(\vec k)$ as basis functions of the irreps, we can now in principle construct all contributions to pairing of any symmetry (irrep). The symmetries of products of $\vec k$-dependent functions and matrices $\sigma_i$ follow from the usual rules for products of representations.

For our purposes, it is sufficient to consider the simplest basis functions, i.e., the ones of lowest angular momentum $l$. Higher-order basis functions just add modulations to the gap functions without changing the symmetry. However, accidental nodes are possible. They can only be obtained from a microscopic and material-specific theory. Recent ARPES experiments \cite{CSK24} indicate that the pairing amplitude goes to zero towards the Weyl points but does not contain further nodes.

The basis functions up to $l=6$, expressed in terms of the polar angle $\phi$ of $\vec k$, are given in Table \ref{tab.fphi}. According to Eqs.\ (\ref{eq.A.1b1}) and (\ref{eq.A.1b2}), only the basis functions for $l=0,2,4,6$ ($l=1,3,5$) can occur in $\psi(\vec k)$ ($\mathbf{d}(\vec k)$). Thus the lowest-order spin-singlet contributions to pairings belonging to the three irreps are
\begin{align}
A_1: && \hat D(\vec k) &\propto \sigma_0 , \label{eq.A.1c1} \\
A_2: && \hat D(\vec k) &\propto \sin6\phi\: \sigma_0 , \\
E: && \hat D(\vec k) &\propto \left\{\begin{matrix}
  \sin2\phi\: \sigma_0 \\[0.5ex]
  \cos2\phi\: \sigma_0 .
\end{matrix}\right. \label{eq.A.1c3}
\end{align}
The lowest-order spin-triplet contributions with spin-components $\sigma_z$ are
\begin{align}
A_1: && \hat D(\vec k) &\propto \cos3\phi\: \sigma_z , \\
A_2: && \hat D(\vec k) &\propto \sin3\phi\: \sigma_z , \\
E: &&  \hat D(\vec k) &\propto \left\{\begin{matrix}
  \sin\phi\: \sigma_z \\[0.5ex]
  -\cos\phi\: \sigma_z ,
\end{matrix}\right.
\end{align}
and the lowest-order spin-triplet contributions with in-plane spin components are
\begin{align}
A_1: && \hat D(\vec k) &\propto -\sin\phi\: \sigma_x + \cos\phi\: \sigma_y , \\
A_2: && \hat D(\vec k) &\propto \cos\phi\: \sigma_x + \sin\phi\: \sigma_y \\
E: && \hat D(\vec k) &\propto \left\{\begin{matrix}
  -\cos\phi\: \sigma_x + \sin\phi\: \sigma_y \\[0.5ex]
  \sin\phi\: \sigma_x + \cos\phi\: \sigma_y .
\end{matrix}\right.
\label{eq.A.1e3}
\end{align}
Generically, all contribution of the same symmetry (irrep) coexist \cite{endnote.purelow}. Note that all pairing matrices in Eqs.\ (\ref{eq.A.1c1})--(\ref{eq.A.1e3}) are even under time reversal, which inverts $\mbox{\boldmath$\sigma$}$ and shifts $\phi \to \phi + \pi$ modulo $2\pi$. They thus belong to the irreps $A_{1+}$, $A_{2+}$, and $E_+$.

Since $\sigma_0$ has full symmetry the $\vec k$-dependent prefactors in Eqs.\ (\ref{eq.A.1c1})--(\ref{eq.A.1c3}) already give the $A_1$, $A_2$, and $E$ pairing amplitudes shown in Fig.\ \ref{fig.basis}. The next step is to show that the triplet contributions lead to momentum dependence of the same symmetry when projected onto the low-energy band.

To obtain the low-energy Hamiltonian, we first perform a unitary transformation into the band basis. The BdG Hamiltonian to the band basis is given by
\begin{equation}
\tilde{\mathcal{H}}(\vec k) \equiv \begin{pmatrix}
  \tilde H_N(\vec k) & \tilde\Delta(\vec k) \\
  \tilde\Delta^\dagger(\vec k) & -\tilde H_N^T(-\vec k)
\end{pmatrix}
= \mathcal{U}(\vec k)\, \mathcal{H}(\vec k)\, \mathcal{U}(\vec k)^\dagger ,
\end{equation}
with
\begin{equation}
\mathcal{U}(\vec k) = \begin{pmatrix}
  \hat U(\vec k) & 0 \\ 0 & \hat U^*(-\vec k)
\end{pmatrix}
= \begin{pmatrix}
  \hat U(\vec k) & 0 \\ 0 & \hat U(\vec k) \hat U_T
\end{pmatrix} ,
\end{equation}
where the final form makes use of the time-reversal symmetry of $\hat H_N$.

In the band basis, the normal-state Hamiltonian is diagonal,
\begin{equation}
\tilde H_N(\vec k) = \begin{pmatrix}
  \xi_-(\vec k) & 0 \\ 0 & \xi_+(\vec k)
\end{pmatrix} ,
\end{equation}
where $\xi_-(\vec k)$ is the dispersion of the band forming the Fermi arcs, without loss of generality. The states with dispersion $\xi_+(\vec k)$ are then high-energy states \cite{endnote.purelow}. The projected BdG Hamiltonian for the low-energy band forming the Fermi arcs is obtained from $\tilde{\mathcal{H}}(\vec k)$ by taking only the matrix elements pertaining to the ``$-$'' band,
\begin{equation}
\mathcal{H}_\mathrm{proj}(\vec k) = \begin{pmatrix}
  \xi_-(\vec k) & \Delta(\vec k) \\
  \Delta^\ast(\vec k) & -\xi_-(-\vec k)
\end{pmatrix} .
\label{eq.A.1f}
\end{equation}
We use $\xi(\vec k) \equiv \xi_-(\vec k)$ in Sec.\ \ref{sec.nodal}. We have checked numerically (not shown) that the triplet contributions indeed lead to momentum dependence of $\Delta(\vec k)$ of the same symmetry as the singlet contributions belonging to the same irrep.

However, we actually do not need to perform this unitary transformation and projection to determine the symmetries of the gap functions $\Delta(\vec k)$, as stated in Sec.\ \ref{sec.nodal}. $\Delta(\vec k)$ is a real function of $\vec k$ (it can be chosen real because the system does not break time-reversal symmetry). It must transform like the pairing matrix $\hat D(\vec k)$ under the point group and under time reversal since the unitary transformation to the band basis must not change the symmetry. This means that $\Delta(\vec k)$ must be a real basis function of the same irrep, i.e., $A_{1+}$, $A_{2+}$, or $E_+$, as $\hat D(\vec k)$. The lowest-order basis functions are the ones given in the manuscript, $1$ for $A_{1+}$, $\sin6\phi$ for $A_{2+}$, and $(\sin2\phi,\cos2\phi)$ for the two components of $E_+$. These functions are plotted in Fig.~\ref{fig.basis}.

\section{Enumeration of full-symmetry products}
\label{app.products}

It is possible to use symbolic algebra to generate all terms with $n$ factors out of $\{A_1, A_2, \E1, \E2\}$ that have full symmetry, i.e., that transform according to $A_1$. For $n=2$, the results are clear,
\begin{align}
&A_1 \otimes A_1 , \\
&A_2 \otimes A_2 , \\
&\E1 \otimes \E1 + \E2 \otimes \E2 .
\end{align}
For $n=3$, the terms are
\begin{align}
&A_1 \otimes A_1 \otimes A_1 , \\
&A_1 \otimes A_2 \otimes A_2 , \\
&A_1 \otimes \E1 \otimes \E1 + A_1 \otimes \E2 \otimes \E2 , \\
&A_2 \otimes A_1 \otimes A_2 , \\
&A_2 \otimes A_2 \otimes A_1 , \\
&A_2 \otimes \E1 \otimes \E2 - A_2 \otimes \E2 \otimes \E1 , \\
&\E1 \otimes A_1 \otimes \E1 + \E2 \otimes A_1 \otimes \E2 , \\
&\E1 \otimes A_2 \otimes \E2 - \E2 \otimes A_2 \otimes \E1 , \\
&\E1 \otimes \E2 \otimes A_2 - \E2 \otimes \E1 \otimes A_2 , \\
&\E1 \otimes \E1 \otimes A_1 + \E2 \otimes \E2 \otimes A_1 , \\
&\E1 \otimes \E1 \otimes \E2 + \E1 \otimes \E2 \otimes \E1 \nonumber \\
&\quad{}+ \E2 \otimes \E1 \otimes \E1 - \E2 \otimes \E2 \otimes \E2 .
\end{align}
For $n=4$, the terms are
\begin{widetext}
\begin{align}
&A_1 \otimes A_1 \otimes A_1 \otimes A_1 , \\
&A_1 \otimes A_1 \otimes A_2 \otimes A_2 , \\
&A_1 \otimes A_1 \otimes \E1 \otimes \E1 + A_1 \otimes A_1 \otimes \E2 \otimes \E2 , \\
&A_1 \otimes A_2 \otimes A_1 \otimes A_2 , \\
&A_1 \otimes A_2 \otimes A_2 \otimes A_1 , \\
&A_1 \otimes A_2 \otimes \E1 \otimes \E2 - A_1 \otimes A_2 \otimes \E2 \otimes \E1 , \\
&A_1 \otimes \E1 \otimes A_1 \otimes \E1 + A_1 \otimes \E2 \otimes A_1 \otimes \E2 , \\
&A_1 \otimes \E1 \otimes A_2 \otimes \E2 - A_1 \otimes \E2 \otimes A_2 \otimes \E1 , \\
&A_1 \otimes \E1 \otimes \E2 \otimes A_2 - A_1 \otimes \E2 \otimes \E1 \otimes A_2 , \\
&A_1 \otimes \E1 \otimes \E1 \otimes A_1 + A_1 \otimes \E2 \otimes \E2 \otimes A_1 , \\
&A_1 \otimes \E1 \otimes \E1 \otimes \E2 + A_1 \otimes \E1 \otimes \E2 \otimes \E1 +  A_1 \otimes \E2 \otimes \E1 \otimes \E1 - A_1 \otimes \E2 \otimes \E2 \otimes \E2 , \\
&A_2 \otimes A_1 \otimes A_1 \otimes A_2 , \\
&A_2 \otimes A_1 \otimes A_2 \otimes A_1 , \\
&A_2 \otimes A_1 \otimes \E1 \otimes \E2 - A_2 \otimes A_1 \otimes \E2 \otimes \E1 , \\
&A_2 \otimes A_2 \otimes A_1 \otimes A_1 , \\
&A_2 \otimes A_2 \otimes A_2 \otimes A_2 , \\
&A_2 \otimes A_2 \otimes \E1 \otimes \E1 + A_2 \otimes A_2 \otimes \E2 \otimes \E2 , \\
&A_2 \otimes \E1 \otimes A_1 \otimes \E2 - A_2 \otimes \E2 \otimes A_1 \otimes \E1 , \\
&A_2 \otimes \E1 \otimes A_2 \otimes \E1 + A_2 \otimes \E2 \otimes A_2 \otimes \E2 , \\
&A_2 \otimes \E1 \otimes \E2 \otimes A_1 - A_2 \otimes \E2 \otimes \E1 \otimes A_1 , \\
&A_2 \otimes \E1 \otimes \E1 \otimes A_2 + A_2 \otimes \E2 \otimes \E2 \otimes A_2 , \\
&A_2 \otimes \E1 \otimes \E1 \otimes \E1 - A_2 \otimes \E1 \otimes \E2 \otimes \E2 - A_2 \otimes \E2 \otimes \E1 \otimes \E2 - A_2 \otimes \E2 \otimes \E2 \otimes \E1 , \\
&\E1 \otimes A_1 \otimes A_1 \otimes \E1 + \E2 \otimes A_1 \otimes A_1 \otimes \E2 , \\
&\E1 \otimes A_1 \otimes A_2 \otimes \E2 - \E2 \otimes A_1 \otimes A_2 \otimes \E1 , \\
&\E1 \otimes A_1 \otimes \E2 \otimes A_2 - \E2 \otimes A_1 \otimes \E1 \otimes A_2 , \\
&\E1 \otimes A_1 \otimes \E1 \otimes A_1 + \E2 \otimes A_1 \otimes \E2 \otimes A_1 , \\
&\E1 \otimes A_1 \otimes \E1 \otimes \E2 + \E1 \otimes A_1 \otimes \E2 \otimes \E1 + \E2 \otimes A_1 \otimes \E1 \otimes \E1 - \E2 \otimes A_1 \otimes \E2 \otimes \E2 , \\
&\E1 \otimes A_2 \otimes A_1 \otimes \E2 - \E2 \otimes A_2 \otimes A_1 \otimes \E1 , \\
&\E1 \otimes A_2 \otimes A_2 \otimes \E1 + \E2 \otimes A_2 \otimes A_2 \otimes \E2 , \\
&\E1 \otimes A_2 \otimes \E2 \otimes A_1 - \E2 \otimes A_2 \otimes \E1 \otimes A_1 , \\
&\E1 \otimes A_2 \otimes \E1 \otimes A_2 + \E2 \otimes A_2 \otimes \E2 \otimes A_2 , \\
&\E1 \otimes A_2 \otimes \E1 \otimes \E1 - \E1 \otimes A_2 \otimes \E2 \otimes \E2 - \E2 \otimes A_2 \otimes \E1 \otimes \E2 - \E2 \otimes A_2 \otimes \E2 \otimes \E1 , \\
&\E1 \otimes \E2 \otimes A_1 \otimes A_2 - \E2 \otimes \E1 \otimes A_1 \otimes A_2 , \\
&\E1 \otimes \E2 \otimes A_2 \otimes A_1 - \E2 \otimes \E1 \otimes A_2 \otimes A_1 , \\
&\E1 \otimes \E2 \otimes \E1 \otimes \E2 - \E1 \otimes \E2 \otimes \E2 \otimes \E1 - \E2 \otimes \E1 \otimes \E1 \otimes \E2 + \E2 \otimes \E1 \otimes \E2 \otimes \E1 , \\
&\E1 \otimes \E1 \otimes A_1 \otimes A_1 + \E2 \otimes \E2 \otimes A_1 \otimes A_1 , \\
&\E1 \otimes \E1 \otimes A_1 \otimes \E2 + \E1 \otimes \E2 \otimes A_1 \otimes \E1 + \E2 \otimes \E1 \otimes A_1 \otimes \E1 - \E2 \otimes \E2 \otimes A_1 \otimes \E2 , \\
&\E1 \otimes \E1 \otimes A_2 \otimes A_2 + \E2 \otimes \E2 \otimes A_2 \otimes A_2 , \\
&\E1 \otimes \E1 \otimes A_2 \otimes \E1 - \E1 \otimes \E2 \otimes A_2 \otimes \E2 - \E2 \otimes \E1 \otimes A_2 \otimes \E2 - \E2 \otimes \E2 \otimes A_2 \otimes \E1 , \\
&\E1 \otimes \E1 \otimes \E1 \otimes A_2 - \E1 \otimes \E2 \otimes \E2 \otimes A_2 - \E2 \otimes \E1 \otimes \E2 \otimes A_2 - \E2 \otimes \E2 \otimes \E1 \otimes A_2 , \\
&\E1 \otimes \E1 \otimes \E2 \otimes A_1 + \E1 \otimes \E2 \otimes \E1 \otimes A_1 + \E2 \otimes \E1 \otimes \E1 \otimes A_1 - \E2 \otimes \E2 \otimes \E2 \otimes A_1 , \\
&\E1 \otimes \E1 \otimes \E1 \otimes \E1 - \E1 \otimes \E1 \otimes \E2 \otimes \E2 + \E1 \otimes \E2 \otimes \E1 \otimes \E2 + \E1 \otimes \E2 \otimes \E2 \otimes \E1 \nonumber \\
&\quad{} + \E2 \otimes \E1 \otimes \E1 \otimes \E2 + \E2 \otimes \E1 \otimes \E2 \otimes \E1 - \E2 \otimes \E2 \otimes \E1 \otimes \E1 + \E2 \otimes \E2 \otimes \E2 \otimes \E2 , \\
&\E1 \otimes \E1 \otimes \E1 \otimes \E1 + \E1 \otimes \E1 \otimes \E2 \otimes \E2 + \E2 \otimes \E2 \otimes \E1 \otimes \E1 + \E2 \otimes \E2 \otimes \E2 \otimes \E2 .
\end{align}
\end{widetext}
Analogously, high-order terms and also terms of $A_2$, $\E1$, or $\E2$ symmetry can be generated.


\begin{thebibliography}{99}

\bibitem{KBR14}M. Kaiser, A. I. Baranov, and M. Ruck, Bi\textsubscript{2}Pt(\textit{hP}9) by Low-Temperature Reduction of Bi\textsubscript{13}Pt\textsubscript{3}I\textsubscript{7}: Reinvestigation of the Crystal Structure and Chemical Bonding Analysis, Z. Anorg.\ Allg.\ Chem.\ \textbf{640}, 2742 (2014).

\bibitem{SKP20}G. Shipunov, I. Kovalchuk, B. R. Piening, V. Labracherie, A. Veyrat, D. Wolf, A. Lubk, S. Subakti, R. Giraud, J. Dufouleur, S. Shokri, F. Caglieris, C. Hess, D. V. Efremov, B. B\"uchner, and S. Aswartham, Polymorphic PtBi\textsubscript{2}: Growth, structure, and superconducting properties, Phys.\ Rev.\ Materials \textbf{4}, 124202 (2020).

\bibitem{VKV24}A. Veyrat, K. Koepernik, L. Veyrat, G. Shipunov, S. Aswartham, J. Qu, A. Kumar, M. Ceccardi, F. Caglieris, N. P\'erez Rodríguez, R. Giraud, B. B\"uchner, J. van den Brink, C. Ortix, and J. Dufouleur, Dissipationless transport signature of topological nodal lines, Nature Commun.\ \textbf{16}, 6711 (2025).

\bibitem{VLB23}A. Veyrat, V. Labracherie, D. L. Bashlakov, F. Caglieris, J. I. Facio, G. Shipunov, T. Charvin, R. Acharya, Y. Naidyuk, R. Giraud, J. van den Brink, B. B\"uchner, C. Hess, S. Aswartham, and J. Dufouleur, Berezinskii--Kosterlitz--Thouless Transition in the Type‑I Weyl Semimetal PtBi\textsubscript{2}, Nano Lett.\ \textbf{23}, 1229 (2023).

\bibitem{KSV24}A. Kuibarov, O. Suvorov, R. Vocaturo, A. Fedorov, R. Lou, L. Merkwitz, V. Voroshnin, J. I. Facio, K. Koepernik, A. Yaresko, G. Shipunov, S. Aswartham, J. van den Brink, B. B\"uchner, and S. Borisenko, Evidence of superconducting Fermi arcs, Nature \textbf{626}, 294 (2024).

\bibitem{VKF24}R. Vocaturo, K. Koepernik, J. I. Facio, C. Timm, I. C. Fulga, O. Janson, and J. van den Brink, Electronic structure of the surface superconducting Weyl semimetal PtBi\textsubscript{2}, Phys.\ Rev.\ B \textbf{110}, 054504 (2024).

\bibitem{endnote.mirrors}For PtBi\textsubscript{2}, the in-plane lattice vectors are parallel to the mirror planes. The in-plane reciprocal-lattice vectors are orthogonal to the in-plane direct-lattice vectors. Hence, the mirror planes are orthogonal to the in-plane reciprocal lattice vectors and thus to the $\Gamma$M lines, but rather contain the $\Gamma$K lines. Note that there are other lattices with the same Brillouin zone for which the mirror planes contain $\Gamma$M. This is important because the basis functions of irreducible representations are different.

\bibitem{HSV24}S. Hoffmann, S. Schimmel, R. Vocaturo, J. Puig, G. Shipunov, O. Janson, S. Aswartham, D. Baumann, B. B\"uchner, J. van den Brink, Y. Fasano, J. I. Facio, and C. Hess, Fermi Arcs Dominating the Electronic Surface Properties of Trigonal PtBi\textsubscript{2}, Adv.\ Phys.\ Res.\ 2400150 (2024).

\bibitem{CSK24}S. Changdar, O. Suvorov, A. Kuibarov, S. Thirupathaiah, G. Shipunov, S. Aswartham, S. Wurmehl, I. Kovalchuk, K. Koepernik, C. Timm, B. Büchner, I. C. Fulga, S. Borisenko, and J. van den Brink, Topological nodal $i$-wave superconductivity in PtBi\textsubscript{2}, arXiv:2507.01774.

\bibitem{ZCC24}J. Zabala, V. F. Correa, F. J. Castro, and P. Pedrazzini, Enhanced weak superconductivity in trigonal $\gamma$-PtBi\textsubscript{2}, J. Phys.: Condens.\ Matter \textbf{36}, 285701 (2024).

\bibitem{WCZ21}J. Wang, X. Chen, Y. Zhou, C. An, Y. Zhou, C. Gu, M. Tian, and Z. Yang, Pressure-induced superconductivity in trigonal layered PtBi\textsubscript{2} with triply degenerate point fermions, Phys.\ Rev.\ B \textbf{103}, 014507 (2021).

\bibitem{BKS22}D. L. Bashlakov, O. E. Kvitnitskaya, G. Shipunov, S. Aswartham, O. D. Feya, D. V. Efremov, B. B\"uchner, and Yu. G. Naidyuk, Electron-phonon interaction and point contact enhanced superconductivity in trigonal PtBi\textsubscript{2}, Low Temp.\ Phys.\ \textbf{48}, 747 (2022).

\bibitem{Ber72}V. L. Berezinskii, Destruction of Long-Range Order in One-Dimensional and Two-Dimensional Systems Possessing a Continuous Symmetry Group. II. Quantum Systems, Sov.\ Phys.\ JETP \textbf{34}, 610 (1972) [Zh.\ Eksp.\ Teor.\ Fiz.\ \textbf{61}, 1144 (1971)].

\bibitem{KoT73a}J. M. Kosterlitz and D. J. Thouless, Ordering, metastability and phase transitions in two-dimensional systems, J. Phys.\ C \textbf{6}, 1181 (1973).

\bibitem{KoT73b}J. M. Kosterlitz, The critical properties of the two-dimensional $xy$ model, J. Phys.\ C \textbf{7}, 1046 (1974).

\bibitem{Tin63a}M. Tinkham, Effect of Fluxoid Quantization on Transitions of Superconducting Films, Phys.\ Rev.\ \textbf{129}, 2413 (1963).

\bibitem{Tin63b}M. Tinkham, \textit{Introduction to Superconductivity}, 2nd edition (Dover, Mineola, 2004).

\bibitem{SFH24}S. Schimmel, Y. Fasano, S. Hoffmann, J. Besproswanny, L. T. Corredor Bohorquez, J. Puig, B.-C. Elshalem, B. Kalisky, G. Shipunov, D. Baumann, S. Aswartham, B. B\"uchner, and C. Hess, Surface superconductivity in the topological Weyl semimetal t-PtBi\textsubscript{2}, Nat.\ Commun.\ \textbf{15}, 9895 (2024).

\bibitem{NoH23}A. Nomani and P. Hosur, Intrinsic surface superconducting instability in type-I Weyl semimetals, Phys.\ Rev.\ B \textbf{108}, 165144 (2023).

\bibitem{TKF24}M. Trama, V. K\"onye, I. C. Fulga, and J. van den Brink, Self-consistent surface superconductivity in time-reversal symmetric Weyl semimetals, arXiv:2410.05381.

\bibitem{DDJ07}M. S. Dresselhaus, G. Dresselhaus, and A. Jorio, \textit{Group Theory: Application to the Physics of Condensed Matter} (Springer, Berlin, 2007).

\bibitem{Katzer}G. Katzer, Character Tables for Point Groups used in Chemistry, http://gernot-katzers-spice-pages.com/character\_tables/index.html.

\bibitem{Gor58}L. P. Gor’kov, On the energy spectrum of superconductors, Sov.\ Phys.\ JETP \textbf{7}, 505 (1958) [Zh.\ Eksp.\ Teor.\ Fiz.\ \textbf{34}, 735 (1958)].

\bibitem{SiU91}M. Sigrist and K. Ueda, Phenomenological theory of unconventional superconductivity, Rev.\ Mod.\ Phys.\ \textbf{63}, 239 (1991).

\bibitem{CTS16}C.-K. Chiu, J. C. Y. Teo, A. P. Schnyder, and S. Ryu, Classification of topological quantum matter with symmetries, Rev.\ Mod.\ Phys.\ \textbf{88}, 035005 (2016).

\bibitem{TiB21}C. Timm and A. Bhattacharya, Symmetry, nodal structure, and Bogoliubov Fermi surfaces for nonlocal pairing, Phys.\ Rev.\ B \textbf{104}, 094529 (2021).


\bibitem{MBT25}K. M\ae{}land, M. Bahari, and B. Trauzettel, Phonon-mediated intrinsic topological superconductivity in Fermi arcs, Phys.\ Rev.\ B \textbf{112}, 104507 (2025).

\bibitem{TsK00}C. C. Tsuei and J. R. Kirtley, Pairing symmetry in cuprate superconductors, Rev.\ Mod.\ Phys.\ \textbf{72}, 969 (2000).

\bibitem{endnote.TR}The sign change stems from $\vec A$, not from $i$. The GL functional maps the fields (OPs and the magnetic field) to the free energy. The functional is invariant under time reversal if it maps the original fields and the time-reversed fields to the same free energy.

\bibitem{MiS94}V. P. Mineev and K. V. Samokhin, Helical phases in superconductors, Sov.\ Phys.\ JETP \textbf{78}, 401 (1994) [Zh.\ Eksp.\ Teor.\ Fiz.\ \textbf{105}, 747 (1994)].

\bibitem{Ede96}V. M. Edelstein, The Ginzburg--Landau equation for superconductors of polar symmetry, J. Phys.: Condens.\ Matter \textbf{8}, 339 (1996).

\bibitem{Agt03}D. F. Agterberg, Novel magnetic field effects in unconventional superconductors, Physica C: Supercond.\ \textbf{387}, 13 (2003).

\bibitem{MiS08}V. P. Mineev and K. V. Samokhin, Nonuniform states in noncentrosymmetric superconductors: Derivation of Lifshitz invariants from microscopic theory, Phys.\ Rev.\ B \textbf{78}, 144503 (2008).

\bibitem{SSY17}M. Smidman, M. B. Salamon, H. Q. Yuan, and D. F. Agterberg, Superconductivity and spin-orbit coupling in non-centrosymmetric materials: a review, Rep.\ Prog.\ Phys.\ \textbf{80}, 036501 (2017).

\bibitem{BaS12}E. Bauer and M. Sigrist, \textit{Non-centrosymmetric Su\-per\-con\-duc\-tors---Introduction and Overview} (Springer, New York, 2012).

\bibitem{endnote.FLE}We have seen above that integration by parts combined with complex conjugation corresponds to the mapping $\psi_i^\ast D_j \psi_k \mapsto - \psi_k^\ast D_j \psi_i$. Noting the explicit minus sign in front of the complex conjugation in Eqs.\ (\ref{disc.A1A1.3})--(\ref{disc.EE.3}), we observe that the superconducting OPs only occur in the combination $\psi_i^\ast D_j \psi_k + \psi_k^\ast D_j \psi_i$. This precludes that the OPs can appear in a combination with $A_2$ symmetry, which would be $\psi_x^\ast D_j \psi_y - \psi_y^\ast  D_j \psi_x$. This reduces the number of possibilities by one.

\bibitem{endnote.energygap}Roughly speaking, the energy gap is proportional to $\sqrt{|\psi_2|^2+|\psi_x|^2}$ since $\psi_2$ and $\psi_x$ have a phase difference of $\pm \pi/2$.

\bibitem{RRJ23}S. H. Ryu, G. Reichenbach, C. M. Jozwiak, A. Bostwick, P. Richter, T. Seyller, and E. Rotenberg, magnetoARPES: Angle Resolved Photoemission Spectroscopy with magnetic field control, J. Electron.\ Spectrosc.\ Relat.\ Phenom.\ \textbf{266}, 147357 (2023).

\bibitem{ADE20}D. F. Agterberg, J. C. S. Davis, S. D. Edkins, E. Fradkin, D. J. Van Harlingen, S. A. Kivelson, P. A. Lee, L. Radzihovsky, J. M. Tranquada, and Y. Wang, The Physics of Pair-Density Waves: Cuprate Superconductors and Beyond, Annu.\ Rev.\ Condens.\ Matter Phys.\ \textbf{11}, 231 (2020).

\bibitem{endnote.purelow}As shown in \cite{VKF24} for the case of $A_1$ pairing, this coexistence allows for pairing to exist only in the low-energy band forming the Fermi arcs and not in the high-energy band of states with reversed spin.

\end{thebibliography}
\end{document}